\documentclass{article}

 \usepackage[dblblindworkshop, final]{neurips_2025}

\usepackage[utf8]{inputenc} 
\usepackage[T1]{fontenc}    
\usepackage{hyperref}       
\usepackage{url}            
\usepackage{booktabs}       
\usepackage{amsfonts}       
\usepackage{nicefrac}       
\usepackage{microtype}      
\usepackage{xcolor}         
\usepackage{graphicx}

\title{The Empty Chair: Using AI Personas to Raise Missing Perspectives in Policy Deliberations}
\workshoptitle{PersonaLLM: Workshop on LLM Persona Modeling}

\author{%
Suyash Fulay \\
Center for Constructive Communication\\
MIT\\
Cambridge, MA \\
\texttt{sfulay@mit.edu
}
\And
Dimitra Dimitrakopoulou \\
Center for Constructive Communication\\
MIT\\
Cambridge, MA \\
\texttt{dimitrad@mit.edu
}
\AND
Deb Roy \\
Center for Constructive Communication\\
MIT\\
Cambridge, MA \\
\texttt{dkroy@mit.edu
}
}

\begin{document}

\maketitle

\begin{abstract}
Deliberation is essential to well-functioning democracies, yet physical, economic, and social barriers often exclude certain groups, reducing representativeness and contributing to issues like group polarization. In this work, we explore the use of large language model (LLM) personas to introduce missing perspectives in policy deliberations. We develop and evaluate a tool that transcribes conversations in real-time and simulates input from relevant but absent stakeholders. We deploy this tool in a 19-person student citizens' assembly on campus sustainability. Participants and facilitators found that the tool was useful to spark new discussions and surfaced valuable perspectives they had not previously considered. However, they also raised skepticism about the ability of LLMs to accurately characterize the perspectives of different groups, especially ones that are already underrepresented. Overall, this case study highlights that while AI personas can usefully surface new perspectives and prompt discussion in deliberative settings, their successful deployment depends on clarifying their limitations and emphasizing that they complement rather than replace genuine participation.
\end{abstract}

\section{Introduction}
Partisan gridlock and increasing political fragmentation have made effective policymaking increasingly difficult \cite{Cohen2003, party_polar}. However, deliberative forums such as citizens' assemblies have shown promise in bypassing party polarization and fostering productive discussions on contentious political issues \cite{guardian_citizens_assembly}. Unfortunately, most deliberations do not take place in carefully structured settings with nationally representative participants. Instead, they often occur within homogeneous groups \cite{mcpherson2001birds}. When this happens, deliberation can lead to group polarization, where individuals become more extreme in their initial positions rather than engaging with opposing viewpoints \cite{sunstein1999law}. This can be problematic if the goal of deliberation is to build common ground and consensus within a pluralistic electorate.
Given that large language models (LLMs) have demonstrated some fidelity in accurately responding to opinion surveys \cite{argyle, park2024generativeagentsimulations1000} and adopting different personas \cite{jiang-etal-2024-personallm}, we explore whether an LLM-powered tool can help introduce missing perspectives in group deliberation. Specifically, we develop and test a system that transcribes discussions in real time and generates input from absent stakeholders. To evaluate its effectiveness, we deploy the tool in small group breakout sessions during a three-day citizens' assembly with university undergraduates, examining whether it encourages participants to engage with perspectives that might otherwise be overlooked. 
From this case study, we had two main findings about the deployment of AI personas in deliberative settings. First, AI personas can help encourage participants to consider perspectives that may be missing from the conversations by making these viewpoints more specific and tangible. However, the way this technology is framed prior to deployment is crucial; presenting personas as a tool to spark new discussions rather than as a replacement for or embodiment of a person or community is more likely to lead to meaningful engagement with said personas.

\section{Background}

\subsection{Perspective-taking to Improve Deliberation}
One promising approach to enhancing deliberation is perspective-taking, defined as the ``active consideration of others' mental states and subjective experiences''. Research has shown that perspective-taking can improve intergroup relations and reduce stereotype expression, among other benefits \cite{Todd2014, Batson1997, Galinsky2000}. Computing systems have also been leveraged to foster perspective-taking. For instance, \citealt{empathosphere} demonstrated that using a bot to regularly prompt team members to consider each other's feelings during a collaboration exercise can enhance communication. \citealt{rpg} show that a computationally supported role-playing game can also induce perspective-taking. Additionally, technology can expose individuals to different subjective experiences, fostering greater understanding and empathy.
In the policy domain, \citealt{crowdsource_perspect} crowd-sourced different perspectives on various policy issues and presented them in an engaging interface, enabling participants to interact with unexpected and differing viewpoints. Similarly, \citealt{help_me} improved deliberation quality by integrating reflection nudges into an online deliberation platform, finding that a persona-based approach was the most effective. This suggests that a persona-driven method could be particularly valuable in live, in-person deliberations, helping participants introduce and consider absent viewpoints in discussions.

\subsection{LLMs as Personas and Representatives}
Since our approach relies on personas to introduce missing perspectives, we briefly survey the literature on the benefits and limitations of using LLMs to model different groups and viewpoints. Several studies have demonstrated that prompted LLMs can reasonably predict survey responses from certain demographic and ideological groups \cite{argyle}. However, they tend to perform worse when modeling minority perspectives \cite{santurkar2023opinionslanguagemodelsreflect} and often risk caricaturing these groups rather than authentically representing them \cite{cheng2023compostcharacterizingevaluatingcaricature}.
Recent work suggests that providing LLMs with significantly more personal context improves their ability to predict individual survey responses and reduce this bias \cite{park2024generativeagentsimulations1000}. Additionally, LLMs have been shown to effectively generate text that aligns with prompted personas, adapting their style and content accordingly \cite{jiang-etal-2024-personallm}. Given the demonstrated effectiveness of personas in prompting reflection to improve deliberation quality \cite{help_me} and the ability of LLMs to model different viewpoints with reasonable accuracy, we investigate whether a tool that dynamically generates personas and their responses in a deliberative setting can effectively help the group raise and understand missing perspectives.
\section{Assembly Information}
Since citizens' assemblies are a key forum for policy deliberation, we deployed our tool in a relatively realistic setting: an in-person, three-day student assembly. The assembly convened 19 undergraduate students to deliberate on what policies or initiatives the university should prioritize to meet its net-zero emissions commitment. This group was heavily biased towards young, educated students interested in sustainability, which was the ideal use case for a tool meant to identify blind spots and raise missing perspectives. To maintain anonymity, we have redacted the university name and any identifying details about participants.
\section{Tool Description}
\begin{figure}
    \centering
    \includegraphics[width=1\linewidth]{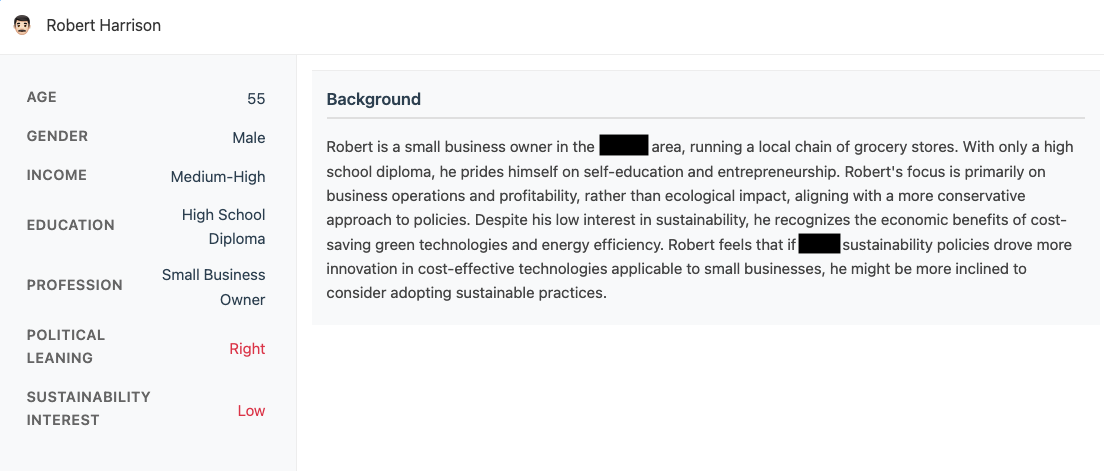}
    \caption{An example persona generated from the conversation transcript and deliberation context. We have redacted any specific references to locations or universities.}
    \label{fig:persona}
\end{figure}

\autoref{fig:flow} outlines the process by which the tool creates dynamic personas that provide relevant input to the conversation. Using a transcript of the discussion along with brief contextual information about the topic and assembly setting, the tool identifies three stakeholder personas who may be affected by the assembly’s decisions.
Each persona is displayed with demographic details—including race, gender, income, occupation, political leaning, and level of interest in sustainability (low, medium, high), as well as background story. We observed that common stakeholder personas included local small business owners and university facilities staff responsible for implementing proposed sustainability initiatives. \autoref{fig:persona} provides an example of a generated persona.
These personas are then integrated with the transcript to surface relevant points of disagreement and highlight missing perspectives within the discussion (see \autoref{fig:reflection} for an example). Additionally, since the tool was deployed during a session where participants were tasked with formulating questions for an expert panel, it also allowed users to generate a question from the perspective of a selected persona, helping to ensure that absent viewpoints were actively considered (see \autoref{fig:question} for an example).

\section{Tool Deployment}
The tool was deployed during a one-hour session of the assembly, where participants were divided into four groups of 4–5 people. The session was structured into two main phases. First, participants discussed the expertise of the panel members and drafted potential questions to ask during the panel. During this time, the tool passively recorded and transcribed the conversation. Then, facilitators provided a brief explanation of the tool’s purpose and functionality. They then used the tool to generate several stakeholder bios (see \autoref{fig:persona}). Delegates took turns reading each persona’s bio before selecting one to explore further.
Once a persona was selected, the facilitator clicked on it, revealing key points of disagreement and missing perspectives associated with that stakeholder (see \autoref{fig:reflection}). Delegates read these perspectives and then engaged in a guided discussion about whether they should take this stakeholder’s concerns into account and, if so, how they might address them.
Finally, the facilitator had the option to generate a question from the perspective of the selected stakeholder (see \autoref{fig:question}). Delegates then discussed whether they should include this question in their list for the panel.
\section{Evaluation}
\subsection{Activity Feedback from Delegates}
To evaluate the tool’s effectiveness, we asked participants to complete a post-activity survey assessing their overall impressions and specific aspects of the tool (see \autoref{fig:likert} for results).
Overall, delegates found the activity engaging and thought-provoking ($m = 5.83$, CI = [5.44, 6.23]) and considered it useful ($m = 5.61$, CI = [5.19, 6.03]). They strongly agreed that the tool helped spark further discussion about missing perspectives ($m = 6.16$, CI = [5.71, 6.60]) and found the system easy to use, with clear outputs ($m = 6.05$, CI = [5.71, 6.39]).
One of the most commonly cited benefits was the tool’s ability to introduce perspectives that might have otherwise been overlooked. As one student put it, \textit{“I liked the bringing in of more perspectives that we hadn't thought of, since our assembly in particular is not representative.”} Another student emphasized how the persona biographies made abstract concerns feel more tangible: \textit{“The persona's background that shared the context behind their position, comments, and questions really helped to shape and represent a sort of real person that has a real problem.”}
Delegates particularly valued the points of disagreement and missing perspectives generated by the tool, with some describing it as a way to surface \textit{“blind spots on demand.”} 

Despite its strengths, the tool also raised concerns about the potential for AI to misrepresent perspectives or create a false sense of representation. One delegate warned that it \textit{“may have the potential to actually stifle diverse voices in citizen's assemblies by acting as a poor substitute for actually bringing in diverse voices.”} This delegate suggested that instead of framing the output as fully realized personas, the tool should \textit{“bring up missing points”} to avoid the misleading impression that AI can perfectly represent all viewpoints, which was different than other delegates who appreciated the relatively detailed backgrounds of the personas.
This tension reflects broader debates in social science research, where LLMs are increasingly used to model human perspectives \cite{Hewitt2024Predicting}. The risks are particularly pronounced given that LLMs tend to be less accurate at predicting the opinions of minority groups \cite{santurkar2023opinionslanguagemodelsreflect} and can introduce biases or caricatures depending on how personas are constructed \cite{gupta2024biasrunsdeepimplicit, cheng2023compostcharacterizingevaluatingcaricature}. Addressing these challenges is critical, and we hope this work highlights both the potential benefits and the real risks of using AI to surface missing perspectives in deliberation.
\subsection{Changes in Empathy and Understanding of Differing Opinions}
Beyond gathering direct feedback on the tool, we sought to evaluate whether introducing missing perspectives through our system increased delegates' empathy and understanding toward those with differing opinions. To assess this, we administered a pre-survey before the assembly to capture baseline attitudes and then provided the same survey shortly after the activity.
As shown in \autoref{fig:prepost}, delegates generally exhibited increased empathy toward individuals with different perspectives on sustainability following the activity ($m_{post} - m_{pre} = 0.63$, $m_{post} = 6.11 \pm 0.42$, $m_{pre} = 5.47 \pm 0.51$). Their empathy also increased specifically toward those who outright disagreed with them ($m_{post} - m_{pre} = 0.68$, $m_{post} = 5.53 \pm 0.48$, $m_{pre} = 4.84 \pm 0.50$).
Additionally, participants were less likely to agree with the statement that individuals who do not prioritize sustainability are actively causing societal harm ($m_{post} - m_{pre} = -0.42$, $m_{post} = 4.89 \pm 0.49$, $m_{pre} = 5.32 \pm 0.56$). They also reported a greater ability to hear, understand, and respect arguments that conflicted with their own beliefs ($m_{post} - m_{pre} = 0.63$, $m_{post} = 6.00 \pm 0.45$, $m_{pre} = 5.37 \pm 0.45$).

\section{Discussion}
In this work, we explored the potential and risks of using AI-generated personas to introduce missing perspectives into small-group policy deliberations. Overall, participants responded positively to AI generated personas, finding the tool useful for expanding their consideration of absent viewpoints and prompting discussions in new directions. Many appreciated how the personas made abstract concerns more concrete and surfaced overlooked perspectives, with facilitators noting that the tool encouraged deeper, more candid engagement. Delegates and facilitators both suggested that having the personas more dynamically respond to the conversation would be an interesting and useful addition to the current tool.
However, two key concerns emerged. First, participants worried that AI-generated personas could misrepresent certain groups, potentially replacing rather than complementing authentic voices. Second, some found the personas’ contributions too generic or redundant, limiting their impact. Facilitators suggested that explicitly framing the tool either as a role-playing aid or as a mechanism for surfacing overlooked perspectives could help clarify its purpose and reduce the chance it would be viewed as replacing participants.
Future iterations of this approach could improve persona specificity and grounding by incorporating real-world data, such as community input or structured datasets, to enhance authenticity and relevance. While AI has the potential to enrich deliberation by introducing different viewpoints, careful design is needed to ensure it serves as a supplement rather than a substitute for genuine representation. We hope this work provides useful insights into the benefits and risks of deploying AI personas in a realistic deliberative settings and helps spark ideas for new applications of LLMs and personas in deliberations.

\bibliographystyle{abbrvnat}
\bibliography{neurips_2025}

\appendix
\section{Limitations}
There are several key limitations to our work. First, our study evaluation was primarily qualitative, with the primary goal being to elicit in-depth feedback in a relatively realistic setting (i.e. as part of a multi-day assembly with live, synchronous conversation). However, we did not run a controlled experiment, and thus future work should validate whether introducing AI personas is more effective than a baseline such as a role-playing activity. Second, our sample of university students may not be representative of how other populations may react to such a tool. Since the participants were relatively knowledgeable about the strengths and weaknesses of AI systems, their reactions may be quite different to those with less knowledge or familiarity with these systems. Finally, we did not solicit feedback from people whom the personas purport to represent (e.g. local business owners and university staff). Thus, although the points raised by the personas seemed plausible and effective at increasing perspective-taking, further validation would be needed to determine if they actually represent people or communities well. 

\section{Ethical Considerations}
There are several risks associated with using AI personas to introduce missing perspectives in deliberation. First, if widely adopted, organizations may rely on AI as a substitute rather than an aid for recruiting participants with different views, potentially reducing efforts to ensure genuine representation. Additionally, AI personas are inherently imperfect and may misrepresent the people or groups they aim to portray. If their role is not clearly framed or their contributions are perceived as low quality, they could inadvertently undermine the perspectives they are meant to highlight, leading participants to view those viewpoints more negatively. Finally, if individuals whose perspectives are being represented have no agency or control over how they are depicted, this could further contribute to feelings of misrepresentation.
\section{Additional Figures}
\begin{figure}
    \centering
    \includegraphics[width=1\linewidth]{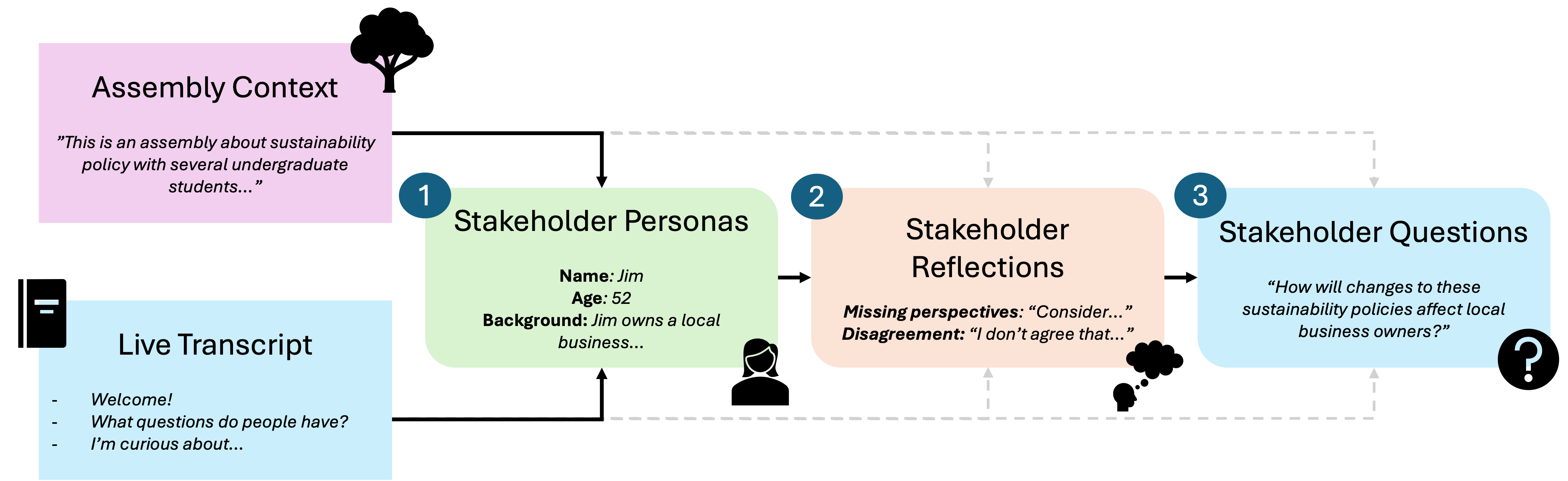}
    \caption{The flow for generating missing stakeholders, stakeholder reflections, and finally stakeholder questions based on the assembly context and a live transcription of the conversation. The solid lines capture how the assembly context and the transcript helps generate missing stakeholder personas, which then is used to generate stakeholder reflections, and then finally the stakeholder questions. }
    \label{fig:flow}
\end{figure}

\begin{figure}
    \centering
    \includegraphics[width=1\linewidth]{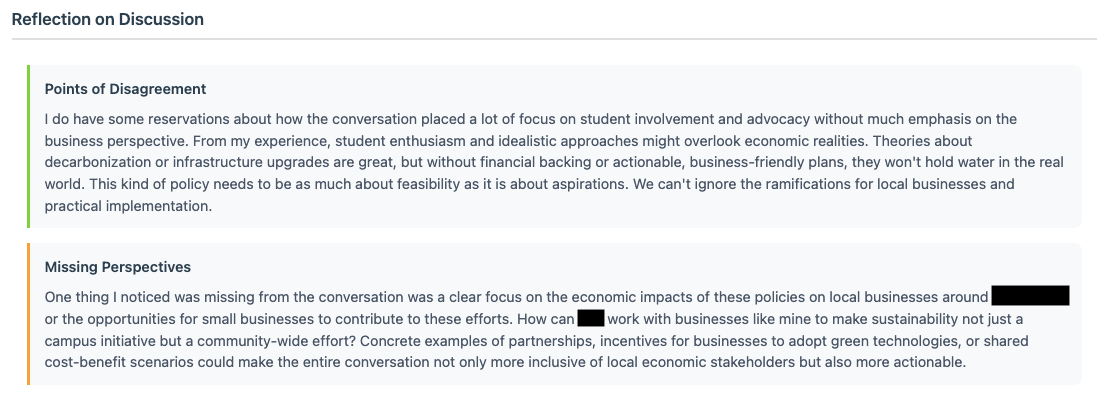}
    \caption{An example of points of disagreement and missing perspectives from a persona.}
    \label{fig:reflection}
\end{figure}

\begin{figure}
    \centering
    \includegraphics[width=1\linewidth]{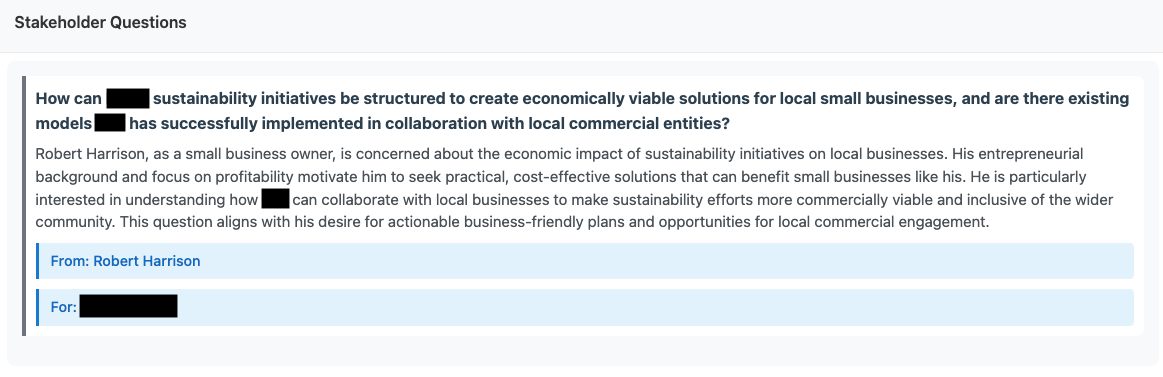}
    \caption{An example question generated by a persona.}
    \label{fig:question}
\end{figure}

\begin{figure}
    \centering
    \includegraphics[width=.9\linewidth]{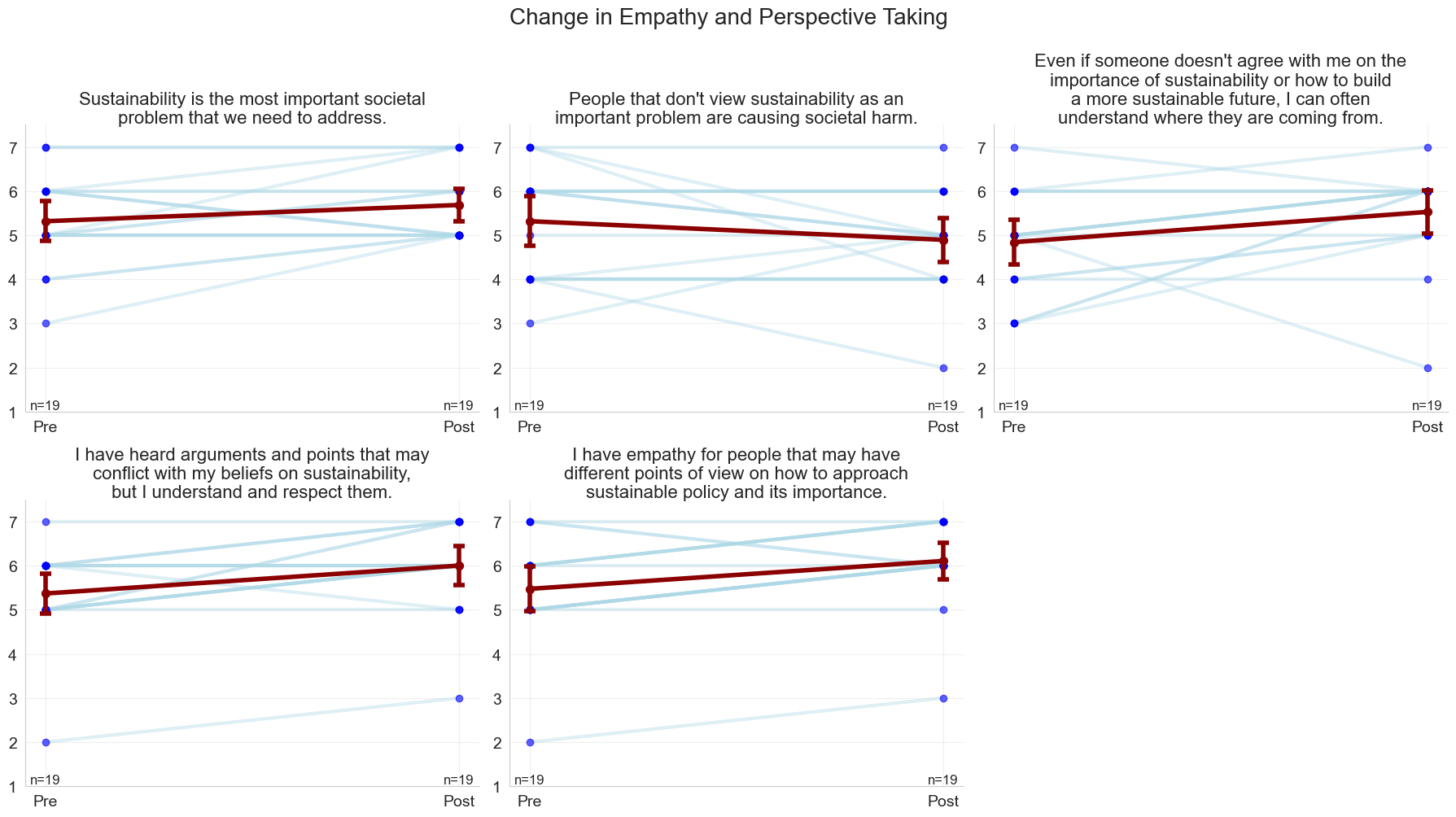}
    \caption{Changes in understanding towards different points between pre-survey (prior to first day) and post-survey (after first day). Red lines capture pre/post means, and blue lines and dots represent individuals.}
    \label{fig:prepost}
\end{figure}

\begin{figure}
    \centering
    \includegraphics[width=1\linewidth]{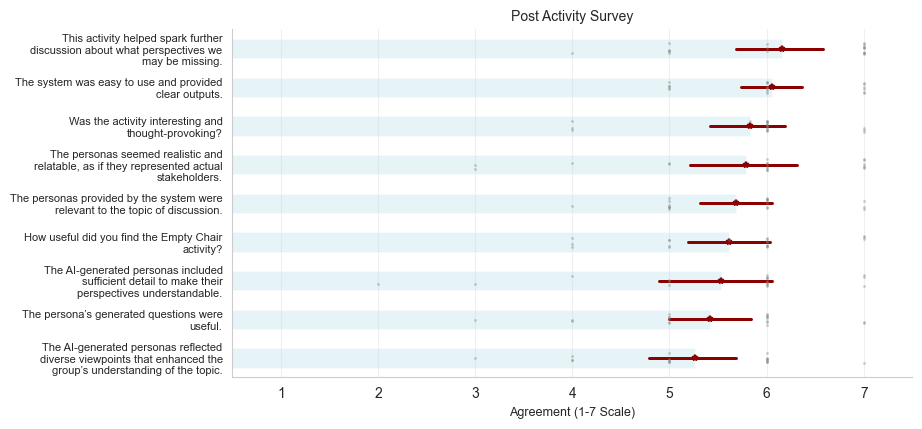}
    \caption{Survey responses capturing how participants felt about various aspects of the activity as well as their overall impressions}
    \label{fig:likert}
\end{figure}

We have added a flow diagram of the context our tool received and what outputs it generated (see \autoref{fig:flow}). We also show the reflection of the AI persona (see \autoref{fig:reflection}) and an example of a generated question (see \autoref{fig:question}). Finally, we plot the average results for our survey responses in \autoref{fig:likert} and \autoref{fig:prepost}.


\section{Technical Details}
For all LLM responses, we used GPT-4o. Given that LLMs may have different capabilities in modeling various perspectives \cite{santurkar2023opinionslanguagemodelsreflect} we hope that in the future tools utilizing AI generated personas have a variety of base model options from which users can choose. Our app was deployed on an AWS EC2 instance that could be accessed from a public URL by facilitators. 
\section{Prompts}
Below we display the system and user prompts for generating stakeholders, stakeholder responses, and stakeholder questions.
\subsection{Stakeholder Generation Prompt}
\textbf{System Prompt:} You are an expert at identifying missing perspectives in discussions. Generate 3 detailed stakeholder bios for people whose perspectives are missing from the conversation. This is a conversation between undergraduate students. Each bio should include:

\begin{itemize}
    \item Name
    \item Description
    \item Demographics:
    \begin{itemize}
        \item Age
        \item Gender
        \item Income
        \item Education
        \item Profession
        \item Political Leaning (Left, Right, Center)
        \item Sustainability Interest (Low, Medium, High)
    \end{itemize}
\end{itemize}

\noindent The response should be formatted as a JSON object with a \texttt{``stakeholders''} array containing objects with \texttt{``name''}, \texttt{``description''}, and \texttt{``demographics''} fields.

\noindent In addition to their thoughts and experience on sustainability, the description should also include personal details about the stakeholder, specifically:

\begin{enumerate}
    \item Their background and experiences
    \item Their beliefs and values
    \item Their personal experiences
    \item Their thoughts on sustainability
    \item Their thoughts on the university's sustainability policies
\end{enumerate}

\noindent Be fairly detailed in the description, including their background, experiences, and beliefs. The demographics should be a JSON object with the specified fields. Make sure to include people who may not be interested or believe in sustainability.

\noindent \textbf{User Prompt:} Based on this transcript, generate 3 missing stakeholders. The discussion, taking place among university undergraduate students, has the following theme:

\noindent\texttt{\{THEME\}}

\noindent The stakeholders should be individuals who may have a vested interest in the university's sustainability policies but are not undergraduate students. For example, they may be other types of university affiliates, people in the area, or individuals who are not affiliated with the university but are still impacted by sustainability policies.
At least one stakeholder should have low sustainability interest.
Format the response as a JSON object with a \texttt{'stakeholders'} array containing objects with \texttt{``name''}, \texttt{``description''}, and \texttt{``demographics''} fields.
The conversation is as follows:

\noindent\texttt{\{transcript\}}
\subsection{Stakeholder Reflections Prompt}

\textbf{System Prompt:}
You are an expert at writing authentic reflections from different perspectives, considering their unique backgrounds, beliefs, and experiences.

\noindent\textbf{User Prompt:} You are an expert at understanding and expressing different perspectives. Given a stakeholder's profile and a conversation transcript, generate a detailed reflection from their perspective. The reflection should feel authentic to their background and personality.
The conversation is part of a broader assembly, the theme of which is discussing sustainability policy at the university.

\noindent{Stakeholder Profile:}
\begin{itemize}
    \item Name: \texttt{\{stakeholder[``name'']\}}
    \item Description: \texttt{\{stakeholder[``description'']\}}
    \item Demographics:
    \begin{verbatim}
    {json.dumps(stakeholder["demographics"], indent=2)\}
    \end{verbatim}
\end{itemize}
The conversation transcript is as follows:
\begin{verbatim}
{transcript}
\end{verbatim}
Generate a reflection from this stakeholder's perspective that includes:
\begin{enumerate}
    \item What they agree with and why (based on their background and experiences)
    \item What they disagree with and why (based on their background and experiences)
    \item What they think is missing from the conversation
\end{enumerate}
\noindent The reflection should be written in the first person and should authentically reflect the stakeholder's:
\begin{itemize}
    \item Education level and profession
    \item Political views (\texttt{\{stakeholder[demographics][political\_leaning]\}})
    \item Interest in sustainability (\texttt{\{stakeholder[demographics][sustainability\_interest]\}})
    \item Personal background and experiences
    \item Socioeconomic status
\end{itemize}
Return the reflection as a JSON object formatted with the following keys:
\begin{itemize}
    \item \texttt{``agree\_explanation''}: Explanation of what they agree with (about 150 words)
    \item \texttt{``disagree\_explanation''}: Explanation of what they disagree with (about 150 words)
    \item \texttt{``missing\_perspectives''}: Explanation of what they think is missing from the conversation (about 150 words)
\end{itemize}
Each explanation should be specific to what was discussed in the conversation. Do not be overly polite or positive. Do not mention demographics explicitly in the reflection.

\section{Generating Stakeholder Questions}
\textbf{System Prompt}: You are an expert at generating insightful questions from different perspectives.Given a stakeholder's profile and a conversation transcript, generate a question that this stakeholder would likely ask, based on their background, demographics, interests, and their reflection on the discussion. The question should be relevant to the topic of university sustainability and should add a new perspective to the conversation. The question should be directed at an expert if possible. Extrapolate the stakeholder's perspective from the transcript and demographics and ask questions someone like this would ask, but don't be obvious about it. Keep the questions relatively general, not too specific to the background of the stakeholder. Keep the questions relatively short and concise.

\noindent\textbf{User Prompt}: 

\noindent Stakeholder Profile:\
\begin{enumerate}
    \item Name: {stakeholder[``name'']}
    \item Description: {stakeholder[``description'']}
    \item Demographics: {json.dumps(stakeholder[``demographics''], indent=2)} 
\end{enumerate}  

\noindent \{reflection\_text\}

\noindent Current Conversation: \{transcript\}

\noindent  Current Questions: \{current\_questions\_text\} 

\noindent The experts are: \{EXPERTS\} 

\noindent Generate a question that this stakeholder would ask, based on their perspective, background, and reflection on the discussion. The question should align with their points of disagreement and what they think is missing from the conversation. Format the response as a JSON object with ``question'',  ``explanation'', and ``expert'' fields, where expert is the expert the question should be directed at. The explanation should justify why this stakeholder would ask this question

\section{Institutional Review Board}
All studies were approved by the Institutional Review Board (IRB) at anonymous institution. We obtained informed consent from all participants and participants were compensated for their time.

\end{document}